\documentstyle[epsfig,12pt]{article}
\begin{document}
\baselineskip=15pt

\begin{center}

{\Huge \bf --------------------------}

{\tiny 

SUNY Institute of Technology at Utica/Rome\\

Conference on Theoretical High Energy Physics\\

June 6th, 2002
}


{\Huge \bf --------------------------}

\end{center}
\vskip 1cm	

\begin{center}
{\Large\bf Optimal RG-Improvement of Perturbative Calculations in QCD}

\vskip .8cm

{\sc Elias, V. $^{a,}$}
\footnote{Electronic address: {\tt velias@uwo.ca}}
\footnote{Permanent address:  {\tt Department of Applied Mathematics, The University of Western Ontario, London, Ontario  N6A 5B7,  Canada}}

\vskip .8cm
{\small \it
$^a$ 
Perimeter Institute for Theoretical Physics, 35 King Street North,
Waterloo, Ontario  N2J 2W9,  Canada}
\\
\end{center}

\vskip 1cm	

\begin{center}
{\Large \bf ------------------------------------------------------------}
\end{center}
\vskip 1cm	

\begin{center}
{\bf  Abstract}\\
\end{center}

{\small \sl
Using renormalization-group methods, differential equations can be obtained
for the all-orders summation of leading and subsequent non-leading 
logarithmic corrections to QCD perturbative series for a number of
processes and correlation functions.  For a QCD perturbative series
known to four orders, such as the $e^+ e^-$ annihilation cross-section,
explicit solutions to these equations are obtained for the summation to
all orders in $\alpha_s$ of the leading set and the subsequent two 
non-leading sets of logarithms.  Such summations are shown for a number of
processes to lead to a substantial reduction in sensitivity to the
renormalization scale parameter.  Surprisingly, such summations are also
shown to lower the infrared singularity within the perturbative
expression for the $e^+ e^-$ annihilation cross-section to coincide with
the Landau pole of the naive one-loop running QCD couplant.
}

\sf

\newpage

\setcounter{footnote}{0}


\section{Optimal RG Improvement of $\Gamma (B \rightarrow X_u \ell^- \bar\nu_\ell)$}

Optimal renormalization-group (RG) improvement of a perturbative series 
to a given order in the expansion couplant is the idea of 
including within that series all higher-order contributions that can be extracted by
renormalization-group methods \cite{maxwell}.  We call such terms, which
involve leading and successive logarithms of the renormalization scale
$\mu$, RG-accessible.  Techniques have been developed to 
obtain closed-form summations of such RG-accessible contributions
to all orders in the perturbative expansion parameter, and such
summations have been shown to lead to ``optimally RG-improved'' 
expressions for perturbative quantities that have
significantly diminished dependence on $\mu$ \cite{ahmady2, ahmady3}.

For example, leading and subleading perturbative QCD corrections to the inclusive
semi-leptonic $B \rightarrow X_u \ell^- \bar\nu_\ell$ decay rate, which
in tree-order is purely a charged-current weak interaction process, are
given by a QCD series \cite{ritbergen}
\begin{eqnarray}
S & = & 1 + \left[ 4.25360 + 5 \log \left( \frac{\mu^2}{m_b^2 (\mu)}
\right) \right] \left( \frac{\alpha_s (\mu)}{\pi} \right)\nonumber\\
& + & \left[ 26.7848 + 36.9902 \log \left( \frac{\mu^2}{m_b^2 (\mu)}
\right) + \frac{415}{24} \log^2 \left( \frac{\mu^2}{m_b^2 (\mu)}
\right) \right] \left( \frac{\alpha_s (\mu)}{\pi} \right)^2 \nonumber\\
& + & {\cal{O}} \left( \left( \frac{\alpha_s (\mu)}{\pi} \right)^3
\right)
\end{eqnarray}
such that
\begin{equation}
\Gamma \left( b \rightarrow u \ell^- \bar\nu_\ell \right) = \frac{G_F^2 |V_{ub}|^2}{192 \pi^3} m_b^5 (\mu) S.
\end{equation}

If one substitutes eq. (1) into eq. (2), one obtains a decay rate that decreases
monotonically with increasing $\mu$, raising the question as to which value of $\mu$ is most appropriate for comparing the calculation (2) to experiment.  
Clearly, such dependence on the unphysical parameter $\mu$ is an embarassment;  indeed the renormalization group equation for the series $S$
\begin{equation}
0 = \left[ \mu^2 \frac{\partial}{\partial \mu^2} + \beta (g) \frac{\partial}{\partial g^2} + m_b \gamma_m (g) \frac{\partial}{\partial m_b} + 5\gamma_m (g) \right] \; \; S\left[ \mu^2, g^2 (\mu), m_b (\mu)\right]
\end{equation}
is nothing more than a chain rule expression for the requirement that the physically measurable decay rate 
be impervious to changes in the renormalization scale parameter $\mu$,
\begin{equation}
0 = \frac{d}{d\mu^2} \Gamma \left( B \rightarrow X_u \ell^-
\bar\nu_\ell \right).
\end{equation}
The residual $\mu$-dependence of the decay rate obtained from the series
(1) is necessarily a consequence of the truncation of that series, as
well as the relatively large value of the expansion constant $\alpha_s
(\mu) / \pi$.

In fact, the series (1) may be expressed as a double summation over
powers of logarithms and the expansion parameter, i.e., in the following
form:
\begin{equation}
S\left[ x, L\right] = \sum_{n=0}^\infty \sum_{m=0}^n T_{n,m} x^n L^m,
\end{equation}
where
\begin{equation}
x \equiv \alpha_s (\mu) / \pi, \; \; \; L \equiv \log(\mu^2 / m_b^2 (\mu)).
\end{equation}
The first few constants of this series, i.e., the set $\{T_{0,0} (=1),
T_{1,0}, T_{1,1}, T_{2,0}, T_{2,1}, T_{2,2}\}$, are given by eq. (1).
However, all higher-order constants of the form $T_{n,n}$, $T_{n, n-1}$
and $T_{n,n-2}$ can be obtained via eq. (3), and, hence, are RG-accessible.
In terms of the new variables $x$ and $L$, the RG-equation (3) may be
expressed as
\begin{equation}
0 = \left[ (1 - 2\gamma_m (x) ) \frac{\partial}{\partial L} + \beta(x)
\frac{\partial}{\partial x} + 5\gamma_m (x) \right] S[x, L].
\end{equation}
If we substitute the series (5) into the RG-equation (7), as well as the 
known QCD series expansions of the RG-functions
\begin{equation}
\beta(x) = -\sum_{n=0}^\infty \beta_n x^{n+2}, \; \; \gamma_m (x) = -
\sum_{n=0}^\infty \gamma_n x^{n+1},
\end{equation}
we find for any integer $p$ that the aggregate coefficients of $x^p
L^{p-1}, \; x^p L^{p-2}$ and $x^p L^{p-3}$ on the right hand side of eq. (7)
necessarily vanish:
\begin{equation}
x^p L^{p-1}: \; \; 0 = p T_{p,p} - \beta_0 T_{p-1, p-1} (p-1) -
5\gamma_0 T_{p-1, p-1}
\end{equation}
\begin{eqnarray}
x^p L^{p-2}: \; \; 0 & = & (p-1) T_{p, p-1} + 2\gamma_0 (p-1) T_{p-1, p-1}
- \beta_0 (p-1) T_{p-1, p-2}\nonumber\\
& - & \beta_1 (p-2) T_{p-2, p-2} - 5\gamma_0 T_{p-1, p-2} - 5\gamma_1 T_{p-2, p-2}
\end{eqnarray}
\begin{eqnarray}
x^p L^{p-3}: \; \; 0 & = & (p-2) T_{p, p-2} + 2\gamma_0 (p-2) T_{p-1, p-2} + 2\gamma_1 (p-2) T_{p-2, p-2}\nonumber\\
& - & \beta_0 (p-1) T_{p-1, p-3} - \beta_1 (p-2) T_{p-2, p-3} - \beta_2 (p-3) T_{p-3, p-3}\nonumber\\
& - & 5\gamma_0 T_{p-1, p-3} - 5\gamma_1 T_{p-2, p-3} - 5\gamma_2 T_{p-
3, p-3} .
\end{eqnarray}  
Given knowledge of $T_{0,0} (=1)$, one can calculate any coefficient
$T_{p,p}$ through successive applications of eq. (9).  Indeed the eq. (1) values $T_{1,1} = 5$ and $T_{2,2} = 415/24$ 
follow from just two successive iterations of (9) using the $n_f = 5$ QCD values $\gamma_0 = 1, \; \beta_0 = 23/12$.  
Similarly, knowledge of all $T_{p,p}$ coefficients, as obtained via (9), plus knowledge of $T_{1,0} = 4.25360$ [eq. (1)]
is sufficient via successive applications of (10) to determine all coefficients $T_{p, p-1}$.  
Finally, knowledge of all coefficients $T_{p, p}$ $T_{p, p-1}$ plus the single coefficient $T_{2,0} = 26.7848$
is sufficient via successive applications of (11) to determine all coefficients $T_{p, p-2}$, since
the set of $\overline{MS}$ RG-function coefficients $\beta_0, \beta_1, \beta_2, \gamma_0 (=1), \gamma_1$, and $\gamma_2\}$
have all been calculated \cite{gross}. 

Since we now see that all coefficients $T_{p,p}, T_{p, p-1}$ and $T_{p,
p-2}$ are RG-accessible, it makes sense to restructure the double-
summation series (4) in the form
\begin{equation}
S[x, L] = \sum_{p=0}^\infty T_{p,p} x^p L^p + \sum_{p=1}^\infty T_{p, p-1} x^p L^{p-1}
+ \sum_{p=2}^\infty T_{p, p-2} x^p L^{p-2} + \sum_{p=3}^\infty T_{p, p-3} x^p L^{p-3} + ... \; ,
\end{equation}
since the first three terms above are completely determined by eqs. (9),
(10) and (11).  We express (12) in the more compact form
\begin{eqnarray}
S[x,L] & = & \sum_{n=0}^\infty x^n \left[ \sum_{p=n}^\infty T_{p, p-n}
(x L)^{p-n} \right]\nonumber\\
& \equiv & \sum_{n=0}^\infty x^n S_n (x L)
\end{eqnarray}
and note that $S_0 (x L)$, $S_1 (x L)$ and $S_2 (x L)$ all correspond to
RG-accessible functions, based upon the information given in (1).
Indeed, the program of optimal RG-improvement is nothing more than the
explicit closed-form evaluation of these functions, and their subsequent
incorporation into the calculated decay rate.

To evaluate the summation $S_0 (u)$, as defined by (13), we simply
multiply eq. (9) by $u^{p-1}$ and sum from $p = 1$ to infinity:
\begin{eqnarray}
0 & = & \sum_{p=1}^\infty p T_{p, p} u^{p-1} - \beta_0 \sum_{p=1}^\infty
(p-1) T_{p-1, p-1} u^{p-1}\nonumber\\
& - & 5\gamma_0 \sum_{p=1}^\infty T_{p-1, p-1} u^{p-1}\nonumber\\
& = & (1 - \beta_0 u ) \frac{d S_0}{du} - 5\gamma_0 S_0 .
\end{eqnarray}
We note from the definition (13) of the series $S_n (x L)$ that
\begin{equation}
S_n (0) = T_{n, 0}.
\end{equation}
The solution of the differential equation (14) with initial condition
$S_0 (0) = T_{0,0} = 1$ is
\begin{equation}
S_0 (u) = (1 - \beta_0 u)^{-5\gamma_0 / \beta_0}.
\end{equation}

A similar procedure is employed to find $S_1$, and $S_2$.  If we
multiply eq. (10) by $u^{p-2}$ and then sum from $p = 2$ to $\infty$, we
find after a little algebra that
\begin{equation}
(1 - \beta_0 u) \frac{dS_1}{du} - (\beta_0 + 5\gamma_0) S_1 = 5\gamma_1
S_0 - (2\gamma_0 - \beta_1 u) \frac{dS_0}{du}.
\end{equation}
Substituting the solution (16) into the right hand side of (17) and
noting that $S_1 (0) = T_{1,0}$, we find that
\begin{eqnarray}
S_1 (u) & = & \frac{5(\gamma_0 \beta_1 / \beta_0 - \gamma_1)/\beta_0}{(1
- \beta_0 u)^{5\gamma_0 / \beta_0}}\nonumber\\
& + & \frac{T_{1,0} - 5 (\gamma_0 \beta_1 / \beta_0 - \gamma_1) /
\beta_0 + [5\gamma_0 (2\gamma_0 - \beta_1 / \beta_0)/\beta_0]\log (1 -
\beta_0 u)}{(1 - \beta_0 u)^{(\beta_0 + 5\gamma_0)/\beta_0}}.
\end{eqnarray}
Similarly, we can multiply eq. (11) by $u^{p-3}$ and then sum from $p =
3$ to $\infty$ to obtain the differential equation
\begin{eqnarray}
(1 - \beta_0 u) \frac{d S_2}{du} & - & (2\beta_0 + 5\gamma_0)
S_2\nonumber\\
& = & (\beta_1 u - 2\gamma_0) \frac{dS_1}{du} + (\beta_1 + 5\gamma_1)
S_1 + (\beta_2 u - 2\gamma_1) \frac{dS_0}{du} + 5\gamma_2 S_0
\end{eqnarray}
whose solution is given by eq. (2.28) of ref. [2]. 

\section{Order-by-Order Elimination of Renormalization Scale Dependence}

If one substitutes solutions for $S_0 (x L)$, $S_1 (x L)$ and $S_2 (x
L)$ into eq. (13), one obtains the following optimally RG-improved
version of the series $S$ \cite{ahmady2}:
\begin{eqnarray}
S & \cong & S_0 (x L) + x S_1 (x L) + x^2 S_2 (x L)\nonumber\\
& = & w^{-60/23} + x \left[ -\frac{18655}{3174} w + 10.1310 +
\frac{1020}{529} \log w\right] w^{-83/23}\nonumber\\
& + & x^2 \left[ 13.2231 w^2 - \left( 47.4897 + \frac{3171350}{279841}
\log w \right) w\right.\nonumber\\
& + & \left. \left( 61.0515 + 25.5973 \log w + \frac{719610}{279841} \log^2
w\right)\right]w^{-106/23}
\end{eqnarray}
with
\begin{equation}
w \equiv 1 - \beta_0 x L
\end{equation}
and with $x$ and $L$ given by eq. (6).

When multiplied by $m_b^5 (\mu)$, this expression has the remarkable 
property of being almost entirely 
independent of $\mu$.  Figure 1 of ref. [2] displays a head to head
comparison of the $\mu$ dependence of eq. (20), and 
the same expression with $S$ given by the known terms of
eq. (1).  For the latter case, $[m_b (\mu)]^5 S$ is seen to decrease
from $\approx 2500$ GeV$^5$ to $\approx 1500$ GeV$^5$ as $\mu$ decreases
from $1.5$ GeV to $9.0$ GeV.  For eq. (20), however, the quantity $[m_p
(\mu)]^5 S$ is seen to be $1816 \pm 6$ GeV$^5$ over the same range of
$\mu$, effectively removing all $\mu$-dependence from the optimally 
RG-improved two-loop calculation.  

Such elimination of renormalization-scale dependence via optimal 
RG-improvement is also upheld for a number of
perturbative expressions, including QCD corrections to the inclusive
semileptonic decay of B-mesons to charmed states $(B \rightarrow X_c \ell^-
\bar\nu_{\ell})$, QCD corrections to Higgs boson decays, the perturbative 
portion of the QCD static potential function, the (Standard-Model)
Higgs-mediated $WW \rightarrow ZZ$ cross section at very high energies,
and QCD sum-rule scalar- and vector-current correlation functions
\cite{ahmady2}. This last example is of particular relevance for QCD corrections to the
benchmark electromagnetic cross-section ratio $R(s) = \sigma(e^+ e^-
\rightarrow {\rm hadrons}) / \sigma(e^+ e^- \rightarrow \mu^+ \mu^-)$.
Such QCD corrections are proportional to the imaginary part of the
vector-current correlation function series, a series which is fully
known to three subleading orders in $\alpha_s$ \cite{gorishny}.  For
five active flavours, we have
\begin{eqnarray}
S & \equiv & 3 R(s) / 11\nonumber\\
& = & 1 + x + \left( 1.40924 + \frac{23}{12} L \right) x^2
+ \left( -12.8046 + 7.81875 L + \frac{529}{144} L^2 \right) x^3 +
...\nonumber\\
\end{eqnarray}
where $x = \alpha_s (\mu) / \pi$, as before, and where $L$ is now the
logarithm
\begin{equation}
L \equiv \log (\mu^2 / s).
\end{equation}

Note that dependence on the physical scale $s$ resides entirely in the
logarithm, and that the all-orders series (22) for $S$, a measurable
quantity, is necessarily impervious to changes in $\mu$.  However,
progressive truncations of (22) introduce progressively larger amounts
of renormalization scale dependence.  For example, if the series $S$ is
truncated after all its known terms, as listed in eq. (22), we find for
$\sqrt{s} = 15$ GeV that to order $x^3$, $S$ increases modestly from
$1.0525$ to $1.0540$ as $\mu$ increases from $7.5$ to $30$ GeV.
\footnote{In all estimates presented here, $\alpha_s (\mu)$ is assumed
to evolve via its known 4-loop order $\beta$-function from $\alpha_s
(M_z) = 0.118$ \cite{pdg}.}  Had we truncated the series (22) following
its ${\cal{O}}(x^2)$ term, we find that such a truncation of $S$ now
decreases from $1.056$ to $1.053$ over the same range of $\mu$, doubling
the magnitude of $\mu$-dependence evident over this range.  Finally, if
we consider only the lowest order correction to unity $\left( S = 1 + x
(\mu) \right)$, we find that $S$ decreases from $1.061$ to $1.045$ as
$\mu$ increases from $7.5$ GeV to $30$ GeV.

Optimal RG-improvement of the known terms of the series (22) has been
shown by the same methods delineated above to lead to the following
expression \cite{ahmady2}:
\begin{eqnarray}
S & = & 1 + x/w + x^2 \left[ 1.49024 - 1.26087 \log w \right] /
w^2\nonumber\\
& + & x^3 \left[ 0.115003 w - 12.9196 - 5.14353 \log w + 1.58979 \log^2
w \right] / w^3 + ...
\end{eqnarray}
where $w$ is given by the definition (21), but with $L$ now given by eq.
(23). Eq. (24) is, of course, really a restructured version of the same
infinite series as eq. (21), and similarly must be independent of $\mu$
when taken to all orders.  However, for the series (24) such imperviousness 
to changes in renormalization scale is now evident on an order-by-order basis.
Truncation of the series (24) after its first nonleading term (i.e., $S
= 1 + x/w$) still provides an expression that exhibits less variation
with $\mu$ than all four known terms of the series (22).  As $\mu$ increases
from $7.5$ to $30$ GeV, we find for $\sqrt{s} = 15$ GeV that $1 + x/w$
decreases from $1.0524$ to $1.0516$.  Similarly, truncation of the
series (24) after its third term leads to a slow decrease from $1.0557$
to $1.0553$, and retention of all four known terms leads to an almost
flat value ($1.05372 \pm 0.00004$) over the same $7.5$ GeV - $30$ GeV
spread in $\mu$.

Consequently, we see that the program of optimal RG-improvement, as
described above, is seen to yield {\it order-by-order} 
perturbation-theory predictions which are almost entirely decoupled from the
particular choice of renormalization scale.

\section{Lowering the Infrared Bound on Perturbative Approximations to
$R(s)$}

The optimally RG-improved series (24) is term-by-term singular when $w
= [1 - \beta_0 x(\mu)$ $\log (\mu^2 / s)]$ is zero.  Since $s$ is the
external momentum scale characterising the physical $e^+ e^-$
annihilation process, we see that the use of (24) is possible only if
\cite{ahmady3}
\begin{equation}
s > \mu^2 \exp \left( - \frac{\pi}{\beta_0 \alpha_s (\mu)} \right).
\end{equation}
It is particularly curious that this bound on $s$ corresponds to the
infrared bound on the naive one-loop ($1L$) running couplant $\left(
x_{1L} = (\alpha_s (\mu))_{1L} / \pi \right)$
\begin{equation}
\mu^2 \frac{d x_{1L}}{d\mu^2} = -\beta_0 x_{1L}^2
\end{equation}
whose solution
\begin{equation}
\alpha_s (\mu) = \frac{\pi}{\beta_0 \log (\mu^2 / \Lambda_{1L}^2)}
\end{equation}
can be inverted as follows to express the $1L$ infrared cut-off in terms
of some reference value of $\alpha_s (\mu)$:
\begin{equation}
\Lambda_{1L}^2 = \mu^2 \exp \left( -\frac{\pi}{\beta_0 \alpha_s (\mu)}
\right).
\end{equation}
Consequently, for a given choice of $\mu$ for which $\alpha_s (\mu)$ is
known (e.g. the value $\alpha_s (m_\tau)$ extracted from $\tau$-decay
experiments), we see that each of the terms in the optimally RG-improved series
(24) diverges as $s$ approaches $\Lambda_{1L}^2$ from above.

The idea that QCD perturbative series break down in the infrared is
hardly new, but the location of this breakdown is usually identified
with an IR-divergence in the {\it all-available-orders} evolution of $\alpha_s
(\mu)$, not the naive 1L Landau pole of eq. (28). To consider the IR
boundary of QCD corrections to $e^+ e^-$ annihilation, we find for
three active flavours that QCD corrections to the $e^+ e^-$ annihilation
cross-section are, as before, obtained from the perturbative series
within the imaginary part of the QCD vector current correlation
function \cite{gorishny}:
\begin{eqnarray}
R(s)/2 & = & 1 + x + \left( 1.63982 + \frac{9}{4} L \right)
x^2\nonumber\\
& + & \left( -10.2839 + 11.3792 L + \frac{81}{16} L^2 \right)
x^3\nonumber\\
& + & ...
\end{eqnarray}
where $x = \alpha_s (\mu)$ and where $L = \log (\mu^2 / s)$.  The
standard phenomenological approach to this series is to first recognize
its all-orders invariance under changes in $\mu$, and then to assume
such invariance applies to truncation of the series after its four known
terms.  This (seldom stated) assumption \cite{yndurain} motivates the
choice $\mu^2 = s$ (i.e. $L = 0$) leading to the usual $n_f = 3$
expression \cite{pdg},
\begin{equation}
R(s) = 2\left[ 1 + x (\sqrt{s}) + 1.63982 \; x^2 (\sqrt{2}) - 10.2839 \; x^3
(\sqrt{s}) + ... \right].
\end{equation}
Such an expansion necessarily falls apart in the infrared when
$x(\sqrt{s})$ becomes large.  Indeed, the large coefficient of $x^3
(\sqrt{s})$ in (30) manifests itself in a sharp drop in $R(s)$ at
$\sqrt{s} \cong 650$ MeV \cite{ahmady3}.

It is interesting to compare the known terms in eq. (30) to the optimally RG-improved
version of the known terms in eq. (29) \cite{ahmady2, ahmady3}:
\begin{eqnarray}
R(s) & = & s \left[ 1 + x (\mu) / w (\mu, s)\right. \nonumber\\
& + & x^2 (\mu) \left( 1.63982 - \frac{16}{9} \log \left( w (\mu , s)
\right) \right) / w^2 (\mu, s)\nonumber\\
& + & x^3 (\mu) \left( -1.31057 w (\mu, s) - 8.97333\right.\nonumber\\
& - & \left. \left. 8.99096 \log \left( w (\mu, s) \right) + 3.16049 \log^2 \left(
w(\mu, s) \right) \right) / w^3 (\mu, s) \right]
\end{eqnarray}
where
\begin{equation}
w(\mu, s) = 1 - \frac{9}{4} x (\mu) \log (\mu^2 / s).
\end{equation}
We first note that $x(\mu)$ and $x(\sqrt{s})$ occurring in eqs. (29),
(30), (31) and (32) are evolved through all known orders of the 
$\beta$-function (8).
Consequently, we are free to assign to eq. (31) an $n_f = 3$ empirical
value for $\mu$ ($\mu = m_\tau$ or, alternatively, $\mu = 1$ GeV) safely
outside the infrared region.  To facilitate comparison of eqs. (30) and (31), we
will assume that $x (\sqrt{s})$ devolves via the full $\beta$-function
from this same initial choice of $\mu$ until, for a sufficiently small
value of $s$, $x(\sqrt{s})$ becomes infinite.  The point here is that
for $n_f = 3$, the first four known terms of the $\beta$-function are
all same-sign:  $\left( \beta_0 x^2 + \beta_1 x^3 + \beta_2 x^4 +
\beta_3 x^5 \right) > \beta_0 x^2$.  Thus for a given value of $x$, the
full $\beta$-function is more negative than the one-loop 
$\beta$-function of eq. (26).  Since the evolution of both equations is
referenced to the same initial value $\mu$, the all-orders couplant
$x(\sqrt{s})$ will diverge at a value of $s$ that is {\it larger} than
$\Lambda_{1L}^2$, the Landau pole of eq. (26).  Hence the series (31)
will probe more deeply into the infrared than the series (30) for $R(s)$.

As an example, consider the running
couplant $x(\sqrt{s})$ obtained via a two-loop $\beta$-function 
$\beta (x) = -\beta_0 x^2 - \beta_1 x^3$.  Solution of the differential
equation $s \;  dx / ds = \beta(x)$ with the initial value $x(\mu)$ yields 
the exact constraint
\begin{equation}
\beta_0 \log \left( \frac{\mu^2}{s} \right) = \frac{1}{x(\mu)} - \frac{1}{x(\sqrt{s})}
+ \frac{\beta_1}{\beta_0} \log \left[ \frac{x(\mu) [x(\sqrt{s})+\beta_0/\beta_1]}{x(\sqrt{s}) [x(\mu)+\beta_0/\beta_1]} \right].
\end{equation}
The two-loop Landau pole $s_{2L}$ occurs when $x(\sqrt{s_{2L}}) \rightarrow \infty$, i.e.,
when
\begin{eqnarray}
 s_{2L} & = & \mu^2 \exp \left( -\frac{1}{\beta_0 x (\mu)} \right)
\left[ 1 + \frac{\beta_0}{\beta_1 x(\mu)}
\right]^{\beta_1/\beta_0^2}\nonumber\\
& = & \Lambda_{1L}^2 \left[ 1 + \frac{\beta_0}{\beta_1 x(\mu)}
\right]^{\beta_1/\beta_0^2}.
\end{eqnarray}
Since $\beta_0$, $\beta_1$ and $x(\mu)$ are all positive, $s_{2L} >
\Lambda_{1L}^2$.  Note that $\beta_2$ and $\beta_3$ \cite{ritbergen2} persist
in being positive.  Consequently, for a given initial value $x(\mu)$, 
the singularity in eq. (31)
that occurs at $w(\mu, s) = 0$ (i.e., at $s = \Lambda_{1L}^2$), continues to precede the
Landau singularity of $x(\sqrt{s})$ characterizing eq. (30).  Thus, the optimally RG-improved
eq. (31) extends the applicability of perturbative QCD to lower values of $s$ than in the 
conventional eq. (30) approach to $R(s)$, as is explicitly shown in Fig. 2 of ref. [3].

Finally, we note that one must distinguish between the infrared limitations on the domain
of perturbative approximations to $R(s)$, and any such limitations on
$R(s)$ itself.  For example, each term within the toy series
$\sum_{n=0}^\infty \left[ -x / (s - \Lambda^2)\right]^n$ diverges at $s
= \Lambda^2$, but the function $(s - \Lambda^2) / (s - \Lambda^2 + x)$
from which this series is extracted, is clearly finite at $s
= \Lambda^2$.  Similarly, the all-orders function $R(s)$, as opposed to
truncations of its series representations, may indeed proceed smoothly
to a finite limit as $s \rightarrow 0$ \cite{howe}.  If such is the
case, the best one can hope for in a perturbative series representation
of $R(s)$ is the deepest possible penetration of that series into the
low-$s$ region.

\vskip 1cm
{\noindent \large\bf Acknowledgments:}\\
 
I am grateful to the Natural Sciences and Engineering Council of Canada
for financial support, and to M. R. Ahmady, F. A. Chishtie, A. H. Fariborz,
N. Fattahi, D. G. C. McKeon, T. N. Sherry, A. Squires and T. G. Steele for their
contributions to much of the research summarized in this talk.

\end{document}